\documentclass[journal]{IEEEtran}
\usepackage{amsmath}
\usepackage{amssymb}
\usepackage{footnote}
\usepackage{graphicx}
\newtheorem{mythm}{Theorem}
\newtheorem{cor}{Corollary}
\newtheorem{prop}{Proposition}
\newtheorem{lem}{Lemma}
\newtheorem{CO}{Critical Observation}
\newtheorem{Obs}{Observation}
\newtheorem{Rem}{Remark}

\begin{document}
\title{On Match Lengths, Zero Entropy and Large Deviations - with Application to Sliding Window Lempel-Ziv Algorithm}
\author{Siddharth Jain* and Rakesh K. Bansal$^{\dagger}$\\
Email: sidjain@caltech.edu and rkb@iitk.ac.in}
\date{}
\maketitle

\begin{abstract}
The Sliding Window Lempel-Ziv (SWLZ) algorithm that makes use of recurrence times and match lengths has been studied from various perspectives in information theory literature. In this paper, we undertake a finer study of these quantities under two different scenarios, i) \emph{zero entropy} sources that are characterized by strong long-term memory, and ii) the processes with weak memory as described through various mixing conditions.

For zero entropy sources, a general statement on match length is obtained. It is used in the proof of almost sure optimality of Fixed Shift Variant of Lempel-Ziv (FSLZ) and SWLZ algorithms given in literature. Through an example of stationary and ergodic processes generated by an irrational rotation we establish that for a window of size  $n_w$, a compression ratio given by $O(\frac{\log n_w}{{n_w}^a})$ where $a$ depends on $n_w$ and approaches 1 as $n_w \rightarrow \infty$, is obtained under the application of FSLZ and SWLZ algorithms. Also, we give a general expression for the compression ratio for a class of stationary and ergodic processes with zero entropy.

Next, we extend the study of Ornstein and Weiss on the asymptotic behavior of the \emph{normalized} version of recurrence times and establish the \emph{large deviation property} (LDP) for a class of mixing processes. Also, an estimator of entropy based on recurrence times is proposed for which large deviation principle is proved for sources satisfying similar mixing conditions.\footnote{*- Department of Electrical Engineering, California Institute of Technology, Pasadena, CA, 91125. \\
$\dagger$ - Department of Electrical Engineering, Indian Institute of Technology, Kanpur, India, 208016. \\ This work was carried out when the first author was at Indian Institute of Technology Kanpur. \\A shorter version of this paper was presented at 2013 IEEE Internaional Syposium on Information Theory, Istanbul, Turkey in the form of two papers\\ i) On Match Lengths and Asymptotic Behavior of Sliding Window Lempel-Ziv Algorithm for Zero Entropy Sequences.\\ ii) On Large Deviation Property of Recurrence Times.\\
Copyright (c) 2014 IEEE. Personal use of this material is permitted.  However, permission to use this material for any other purposes must be obtained from the IEEE by sending a request to pubs-permissions@ieee.org.}

\end{abstract}
\section{Introduction}\label{c31}
\subsection{SWLZ algorithm and Zero Entropy Processes}
The analysis of LZ-77 algorithm in probabilistic setting was initiated by Wyner and Ziv~\cite{Ziv89} using the results from ergodic theory. This work was followed by the results on return times for stationary and ergodic sources by Ornstein and Weiss~\cite{Ornstein93}. Further connections between return times and data compression have been obtained by Szpankowski~\cite{S} and Kontoyiannis~\cite{Kontoyiannis98}. Kim and Park~\cite{Kim} have given results on recurrence times for \emph{zero entropy} ergodic processes generated by an irrational rotation.

The Sliding Window Lempel Ziv (SWLZ) algorithm, which is very similar to LZ-77 algorithm, was introduced by Wyner and Ziv~\cite{Ziv94} and its asymptotic optimality for the class of stationary and ergodic sources was proved in expected sense. Shields~\cite{Shields99} proved the  optimality of SWLZ algorithm for individual sequences by comparing the number of phrases generated with that in LZ-78 algorithm. Thus, Shields' results cover almost sure optimality of the SWLZ algorithm for all stationary and ergodic processes just like the optimality of LZ-78. In this paper, we have analysed the behavior of the SWLZ algorithm for stationary and ergodic zero entropy sources. More precisely, we first consider the convergence rate of the compression ratio of the SWLZ algorithm for zero entropy processes generated by irrational rotation and then give a general result over a class of stationary and ergodic zero entropy sources.

Jacob and Bansal~\cite{Bansal} proved the asymptotic optimality of the SWLZ algorithm in almost sure sense by modifying the technique used by Wyner and Ziv~\cite{Ziv94}. Here, following the method used by Jacob and Bansal, we have obtained faster convergence rates for SWLZ algorithm for zero entropy sources as compared to that for positive entropy. However, it is true that the match length $\mathcal{L}_o$ used by Wyner and Ziv as well as Jacob and Bansal can be used to prove the almost sure optimality of SWLZ algorithm for zero entropy processes but in this paper, motivated by the behavior of SWLZ algorithm on periodic sequences (which is discussed in detail in section \ref{c32}) and recent results by Kim and Park~\cite{Kim}\cite{KimD} on recurrence properties for irrational rotations and in general for almost every interval exchange map with zero entropy~\cite{Kim1}, we show that for zero entropy processes it is possible to choose a larger match length $\mathcal{L}_o$ in the proof. For irrational rotations it is very close to the window size as is evident through Eq.(\ref{7}) later. Through this we establish a better convergence rate of the compression ratio that is achievable when SWLZ algorithm is applied to \emph{zero entropy} sequences.

Paul Shields in~\cite{Shields93} showed that there is no universal redundancy rate for any sequence of prefix codes for the class of all ergodic sources. Following this paper in~\cite{Shields95}, Shields and Weiss constructed a class of B-processes for which universal redundancy rates do not exist. Savari ~\cite{Savari98} obtained an upper bound on redundancy of LZ-77 algorithm for unifilar Markov Source. Lastras-Monta$\tilde{n}$o~\cite{Montano06} proved the almost sure optimality of the SWLZ algorithm for the class of processes with exponential rates of entropy using the ideas of Wyner and Ziv~\cite{Ziv94}. Further, he showed that for positive entropy, finite order irreducible markov sources, SWLZ algorithm achieves a compression ratio upper bounded by $H+O(\sqrt{\frac{\log\log n_w}{\log n_w}})$ and lower bounded by $ H + (H+o(1))\frac{\log \log n_w}{\log n_w}$
 for a window of size $n_w$. In this paper, we obtain a finer statement on the upper bound for zero entropy cases. This is discussed in detail in section \ref{c33} and \ref{c34}.

Before we move further, we discuss some points on zero entropy transformations. Zero entropy transformations, defined in section \ref{c33} and giving rise to zero entropy processes, lie in two categories i.e., the one with a discrete spectrum (defined in~\cite{Walters}) and the others with a quasidiscrete (introduced by Abramov~\cite{Abramov}) spectrum. Kushnirenko~\cite{Kushnirenko} proved that a zero entropy transformation is discrete iff for every sequence, the sequence entropy~\cite{Walters} of the transformation is zero. Later, for the transformations with quasidiscrete spectrum, Hulse~\cite{Hulse} gave some information on the sequences for the which sequence entropy is positive.

Entropy dimension as defined in Section 3 is yet another way of classifying zero entropy systems. 
Furthermore, invertible measure preserving transformations on a probability space with zero entropy constitute a dense $G_{\delta}$ (a countable intersection of open sets in weak topology) and therefore outnumber the strong mixing transformations which constitute a set of category one (see page 117 in \cite{Walters}). With the aid of a finite partition they induce a rich class of stationary processes with finite alphabet and zero entropy rate which then become of interest while studying the problem of universal compression algorithms based on recurrence times or match lengths as defined in Section 3. One expects smaller recurrence times and larger match lengths in contrast to positive entropy case where recurrence times grow exponentially and entropy rate is the exponent. This phenomenon translates into faster convergence of the compression ratio to its limit, that is zero, when string matching algorithms are used. Apart from irrational rotation, a second order real valued process, with singular covariance function has zero entropy (see Theorem 16 of Chapter 5 in \cite{Parry}).

In the context of zero entropy processes, in section \ref{c32} we provide our basic motivation towards carrying out a finer analysis of the convergence rate of SWLZ algorithm for zero entropy processes. In section \ref{c33}, we give results on recurrence times and match lengths and establish a theorem on match length which is analogous to corollary \ref{OWL} set up by Ornstein and Weiss~\cite{Ornstein93}. In subsection \ref{c34FDFS-LZ}, we illustrate the compression ratio achievable by Fixed Database Fixed Shift Lempel-Ziv (FDFS-LZ) algorithm for stationary and ergodic sources generated by an irrational rotation. Next in subsections \ref{c34FSLZ} and \ref{c34SWLZ}, we consider Fixed Shift Lempel-Ziv (FSLZ) and SWLZ algorithms respectively. We make a different choice of match length $\mathcal{L}_o$ for zero entropy processes as compared to the one chosen by Wyner and Ziv in~\cite{Ziv94} and Jacob and Bansal in~\cite{Bansal} in their proofs of optimality of these algorithms. Using this $\mathcal{L}_o$, we give a result on convergence rate of the compression ratio that is achievable by applying FSLZ and SWLZ algorithms for a class of zero entropy processes.

\subsection{Large Deviations and Recurrence Times}
 For a stationary and ergodic source with finite alphabet, the asymptotic relationship between probability of an $n$ length sequence and entropy has been well established by Shannon-McMillan-Breiman Theorem~\cite{Shields96}. Later, Ornstein and Weiss~\cite{Ornstein93} established a similar expression relating recurrence times to entropy. Kontoyiannis~\cite{Kontoyiannis98} related recurrence times and probability of an $n$ length sequence for Markov sources by showing that \\ $\lim_{n\rightarrow \infty} \log[R_n(X)\mu(X_1^n)] = o(n^\beta)~a.s.$, for any $\beta > 0$. Here, $\mu(X_1^n)$ is the probability of the $n$ long block $X_1^n$ and $R_n(X)$ is the time it takes to reappear for the first time in $X$, called the recurrence time. $X$ stands for the infinite sequence$..., X_1, X_2,X_3,....$. Further, in \cite[Corollary 2]{Kontoyiannis98}, he also identified a class of processes for which central limit theorem (CLT) and law of iterated logarithm (LIL) hold for recurrence times.

The question of large deviations in Shannon-Mcmillan-Breiman Theorem has been successfully answered in literature under certain mixing conditions~\cite{Shields96}. Motivated by Kontoyiannis' results and the satisfaction of large deviation property for Shannon-McMillan-Breiman Theorem, it is natural to ask under what conditions the asymptotic recurrence times relation satisfies the large deviation property. Chazottes and Ugalde~\cite{Chazottes2005} in 2005 established partial large deviations results on recurrence times for Gibbsian sources. In this paper, we display a class of processes for which large deviation property holds completely for recurrence times.

For an i.i.d source, Shannon-McMillan-Breiman Theorem satisfies large deviation property (LDP) by direct application of Cramer's Theorem~\cite{Dembo}. However for Ornstein and Weiss' result on recurrence times, even for an i.i.d. source Cramer's Theorem is not applicable. This makes the analysis of large deviation property for recurrence times non-trivial even for the i.i.d case. Hence, in order to answer the question of large deviations for recurrence times one needs to look more closely into the recurrence time statistics.

Maurer~\cite{Maurer92} studied the behavior of recurrence time statistics under the assumption of non-overlapping recurrence blocks for i.i.d sources. Later, Abadi and Galves~\cite{AbadiG04} studied a similar non-overlapping scenario for $\psi$-mixing processes and established an exponential bound on the recurrence time distribution. Moreover, they also brought out the contrast between overlapping and non-overlapping case. In the context of overlapping $R_n(x)$, there are several references that show convergence in distribution of $R_n(x)\mu(x_1^n)$ to an exponentially distributed random variable for a fixed $n$-length block $x_1^n$ for a certain class of stationary and ergodic processes~\cite{Abadi04}\cite{Collet99}\cite{Galves97}\cite{Hirata99}. Further, Kim~\cite{Kim12} studied the behavior of conditional distribution of $R_n(X)\mu(X_1^n)$ given the $n$-length block $X_1^n = x_1^n$ and established an exponential bound on its probability for two classes of sources i) $\psi$-mixing, ii) $\phi$-mixing with summable coefficients. In order to carry out our analysis we have used this exponential bound on conditional distribution established by Kim~\cite{Kim12}. We use this bound because it is the only result in our knowledge that establishes bounds on \emph{conditional} probability of recurrence times unlike of others that find bounds on the unconditional probability of recurrence times (where $x_1^n$ is a fixed block). This can be further used to prove LDP for $R_n(X)$, which is the recurrence time for a \emph{random} block $X_1^n$.

 In the context of large deviations and recurrence times,  in section \ref{c43}, we state preliminary results on recurrence time statistics and mixing processes. In section \ref{c44}, we state our main theorems for the large deviation property of recurrence times. In section \ref{c45}, we give proofs of these theorems and their corollaries. In section \ref{c46}, we define an estimator for entropy based on recurrence times and prove large deviation property for it. In section \ref{c47}, we present our conclusion.

\section{Motivation}\label{c32}
Consider two infinite sequences $$x =010101...,$$ $$Tx = 101010......,$$  Here $T$ is the shift transformation, i.e., $(Tx)_{i+1} = x_i$. Let $\mu(x) = \mu(Tx) = 0.5$ be the probability of occurence of $x$ and $Tx$ respectively. So, $\{x,Tx\}$ define the two realizations of a stationary ergodic process which are periodic and entropy rate of the process is zero.

If LZ-78~\cite{Ziv78}\cite{Cover06} is applied on the sequence $x= 0101010...$ then the phrases are generated as follows:
\begin{enumerate}
\item \emph{Initialization:} The initial phrase is the first symbol of the sequence $x$.
\item \emph{Matching:} At step $i$ after initialization step, let $x_1^{L_i}$ be the segment of the sequence until which the phrases have been generated. The next phrase starting from $x_{L_i+1}$ is the minimum length string  such that all but the last symbol in this string together constitute a previously seen phrase.
\end{enumerate}
The \emph{Matching} step is repeated until the sequence $x$ is exhausted.
Hence, for $x = 0101010101...$, the initial phrase = $0$ (phrase length ($l_p$)= $1$). \\
At Step 1 after initialization, $L_1= 1$ and the new phrase generated is $1$ ($l_p = 1$). \\
At Step 2, $L_2 = 2$ and the new phrase generated is $01$ ($l_p = 2$).\\
At Step 3, $L_3 = 4$ and the new phrase generated is $010$ ($l_p = 3$).\\
At Step 4, $L_4 = 7$ and the new phrase generated is $10$ ($l_p= 2$).\\
At Step 5, $L_5 = 9$ and the new phrase generated is $101$ ($l_p = 3$).\\
Following this procedure, the LZ-78 parsing of $x = 010101010...$ is given by:
 $$0,1,01,010,10,101,0101,01010,1010,10101,...$$
 Thus, the total number of phrases generated by applying this procedure is $c(N) \sim N^{\frac{1}{2}}$ for an $N$ long segment of the sequence $x$. Since there are $c(N)$ phrases, therefore $\log c(N)$ bits are required to specify the index of each phrase and an additional one bit is required to specify the last bit of each phrase. Therefore, total number of bits required to encode the sequence $x_1^N$ or in other words all the $c(N)$ phrases $= c(N)(\log c(N) + 1)$,
  which gives the compression ratio to be $$\frac{c(N)\log c(N)+c(N)}{N}  \sim \frac{N^{\frac{1}{2}}\log N +N^{\frac{1}{2}}}{N} \sim \frac{\log N}{N^{\frac{1}{2}}}.$$ However, if SWLZ algorithm as described in subsection \ref{c34SWLZ} is applied on the N-long segment of sequence $x$, $1$ bit will be required to specify the start in the initial window of size $2$ and about $\log N$ bits will be required using alias code to specify the length of the match, since in one iteration complete match will be found in the initial window itself. So, in this case the compression ratio will be $\frac{\log N}{N}$, which converges faster as compared to the compression ratio in the case of LZ-78 applied on N- long segment of the sequence $x$. But for certain positive entropy sources, it has been proved that LZ-78 algorithm achieves faster redundancy rates~\cite{Ziv78}\cite{Kieffer99}\cite{Kieffer2005}\cite{Montano06} as compared to SWLZ algorithm. This fact combined with the above example of a periodic sequence motivates us to study the behavior of SWLZ algorithm on aperiodic sequences generated by a stationary and  ergodic process of \emph{zero} entropy rate.

\section{Recurrence Times and Match Lengths}\label{c33}
Let the sequence $x= \{x_n\}_{-\infty}^\infty$ be an instance of an ergodic source $X= \{X_n\}_{-\infty}^\infty$ with finite alphabet $A$. Here, $x_k^j$ denotes the segment $x_k,x_{k+1},......,x_j$ for sequence $x$. For $n > 0$, define
$$ R_n(x) = \min\{l: l > 0, x_{1}^{n} = x_{-l+1}^{-l+n}\}. $$
Thus, $R_n(x)$ is the first return time of the word $x_{1}^{n}$ in the past.

$$ L_m(x) = \max\{j: j > 0, x_1^j = x_{-k+1}^{-k+j}, k = 1,2,....,m\}.$$
 Therefore, $L_m(x)$ is the length of the longest possible match in the previous $m$ symbols.
\begin{Obs}\label{Obs41}{\textrm{\textit{\cite{Kontoyiannis98}}} $R_n(x) > m \Leftrightarrow L_m(x) < n. $}\end{Obs}
 On the asymptotic behavior of $R_n(x)$ and $L_m(x)$ we have the following,

\begin{mythm}\label{OW}
{\textit{Ornstein and Weiss~\cite{Ornstein93}}\\
With probability 1, for a  stationary and ergodic source $X$
 $$\lim_{n \rightarrow \infty} \frac{\log R_n(x)}{n} = H(X).$$
 where $H(X)$ is the entropy of the source X.
 }
\end{mythm}

\begin{cor}\label{OWL}
{
It has been mentioned in~\cite{Ornstein93} as a corollary of Theorem \ref{OW} that for any stationary and ergodic process with probability 1
$$\lim_{m \rightarrow \infty} \frac{\log m}{ L_m(x)} = H(X). $$
}
\end{cor}
As a general case of corollary \ref{OWL}, consider a stationary and ergodic process for which the asymptotics of the first return time follow
\begin{equation}\label{106}
\lim_{n \rightarrow \infty} \frac{\log R_n(x)}{g(n)} = c.
\end{equation}
where $g(n)$ is an increasing function of $n$ such that $\lim_{n \rightarrow \infty} \frac{g(n+1)}{g(n)} = 1 $, and $ c~\epsilon~(0,\infty)$ is a constant.

\begin{mythm}\label{OW2}
{ With probability 1 for a stationary and ergodic process satisfying Eq. (\ref{106})
$$\lim_{m \rightarrow \infty} \frac{\log m}{g(L_m(x))} = c.$$
}
\end{mythm}
\textit{Proof of Theorem \ref{OW2}:}
Let $n$ and $m$ be natural numbers such that
\begin{equation}\label{107}
n = L_m(x).
\end{equation}
Hence, by definition of a match length we have
\begin{equation}\label{Inequality}
R_{n+1}(x) > m ~and~ R_n(x) \leq m.
\end{equation}

Also we have the following observation:
\begin{Obs}\label{O1}
{ As $n \rightarrow \infty$, $L_m(x) \rightarrow \infty$ and as $L_m(x) \rightarrow \infty$, $m \rightarrow \infty$, by the definition of match length.
}
\end{Obs}
Therefore, using Eq. (\ref{107}) and Inequality (\ref{Inequality}) we have
\begin{equation}\label{Sandwich}
\frac{\log R_{n+1}(x)}{g(n)} > \frac{\log m}{g(L_m(x))} \geq \frac{\log R_n(x)}{g(n)}.
\end{equation}
Now we apply the limit $ n \rightarrow \infty$ on inequality (\ref{Sandwich}) and hence using Sandwich theorem and Observation \ref{O1} we have obtained the statement of Theorem \ref{OW2}.

Now we shift our attention to zero entropy sources. For this purpose, consider the following, let $v ~\in ~\mathcal{R}$ and let $\|v\|$ denote its distance to the nearest integer, i.e.,
$$ \|v\| = \min_{n~\in~ Z} |v-n|. $$

Let $\theta$ be an irrational number in $(0,1)$. Consider $T : [0,1) \rightarrow [0,1)$ to be an irrational rotation by $\theta$. i.e., $$ T(x) = (x+\theta) ~mod ~1. $$
$T$ preserves the Lebesgue measure on $X = [0,1)$.

Given a partition $\mathcal{P}$ and a transformation $T$, in literature the entropy of T with respect to $\mathcal{P}$ is defined as $$ h(\mathcal{P}, T) = \lim_{k\rightarrow \infty}\frac{1}{k} H(\bigvee_{i = 0}^{k-1} T^{-i} \mathcal{P}),$$ and the entropy of the transformation $T$ is $h(T)\triangleq\sup_{\mathcal{P}}h(\mathcal{P},T)$. Here, if $\mathcal{P}= \{P_1,P_2,...P_a\}$ and $\mathcal{Q}= \{Q_1,Q_2,....Q_b\}$, then $\mathcal{P}\bigvee \mathcal{Q} = \{P_i \bigcap Q_j: 1 \leq i \leq a, 1 \leq j \leq b\}$. An irrational rotation in particular, has zero entropy~\cite{Petersen}.

To measure the complexity of entropy zero systems, the notion of entropy dimension $\alpha(\mathcal{P})$ is defined in literature. Let $(X,\mathcal{B},\mu)$ be a probability space and $T$ be a $\mu$ preserving transformation. $(X,\mathcal{B},\mu,T)$ has entropy dimension $\alpha$ with respect to a partition $\mathcal{P}$ of $X$ if $$ \alpha(\mathcal{P}) = \inf \{\beta : \lim_{n\rightarrow \infty} \frac{1}{n^\beta} H(\bigvee_{i = 0}^{n-1} T^{-i} \mathcal{P}) = 0\}, $$ where $H(\mathcal{P})$ is the entropy of the partition $\mathcal{P}$. For a system having positive entropy with respect to $\mathcal{P}$, $\alpha(\mathcal{P}) = 1$~\cite{Kim}. Moreover, for each $0 \leq \alpha \leq 1$, there exists a system of entropy dimension $\alpha$~\cite{Ferenczi}.  For zero entropy systems, $\alpha  < 1$, which means that the number of typical sequences in these systems are of the order of $2^{cn^\alpha}$ (when $0< \alpha < 1$), where $c > 0$. Cheng and Li \cite{Cheng} (Page 1018, Section 3.3, Example 2), point to an example in \cite{Cassaigne} where number of distinct words of length $n$ are about $2^{n^\alpha}$ as $n$ grows to infinity.

Now, we shift our attention to the sequences generated by an irrational rotation. A sequence $x$ made up of zeros and ones is called \emph{Sturmian} if $p_x(n)  = n+1$, where $p_x(n)$ denotes the number of different words of length $n$ occuring in $x$. A necessary and sufficient condition for a sequence $x = x_1x_2x_3..$ to be \emph{Sturmian} is that it is a $\mathcal{P}$ trajectory of an irrational rotation~\cite{Lothaire} which means that if $\mathcal{P} = \{E, E^c\}$ we have $x_i = 0$ if $T^{i-1}x ~\in~ E$ , $1$ otherwise  $\forall ~i = 1,2,3....$. Here $|E| = \theta$ and $|E^c| = 1-\theta$. For other $E$, $p_x(n) = 2n$. 

For an irrational $\theta ~ \in ~(0,1)$, define
$$ \eta(\theta) = \sup \{ t>0 : \liminf_{i\rightarrow\infty} i^t\|i\theta\| = 0\}.$$
In general, $\eta(\theta) \geq 1$ and $= 1$ for almost every $\theta$~\cite{Kim}.

\begin{mythm}\label{KP}
{\textit{Kim and Park~\cite{Kim}}\\
For almost every sequence generated by rotation by an irrational $\theta ~\in ~(0,1)$ :
\begin{equation}\label{1}
\liminf_{n\rightarrow\infty} \frac{\log R_n(x)}{\log n} = \frac{1}{\eta(\theta)}.
\end{equation}
\begin{equation}\label{2}
\limsup_{n\rightarrow\infty} \frac{\log R_n(x)}{\log n} = 1.
\end{equation}
which implies that if $\theta$ is such that $\eta(\theta) = 1$ then the limit exists and is given by:
\begin{equation}\label{3}
\lim_{n\rightarrow\infty} \frac{\log R_n(x)}{\log n} = 1.
\end{equation}
For every other $\theta$ such that $\eta(\theta) > 1$, we have
\begin{equation}\label{4}
\liminf_{n\rightarrow\infty} \frac{\log R_n(x)}{\log n} < 1.
\end{equation}
}
\end{mythm}
Now, applying Theorem \ref{OW2} by choosing the function $g(n) = \log n$, we immediately deduce the following corollary for the case when $\eta(\theta) = 1$

\begin{cor}\label{OWLT}
 { For a stationary and ergodic process generated by irrational rotation where $\eta(\theta) = 1$  with probability 1 $$ \lim_{n \rightarrow \infty} \frac{\log n}{\log L_n(x)} = 1. $$
 }
 \end{cor}

\begin{CO}\label{CO1}
{Consider a stationary and ergodic process with positive entropy. By Theorem \ref{OW} and Egoroff's Theorem~\cite{Royden68}, given $\delta > 0$ and $\epsilon > 0$, $\exists$ a set $F_1$ of sequences $x$ with $\mu(F_1) > 1-\delta$ and an integer $N_o(\epsilon,F_1)$ s.t.
$$ H-\epsilon < \frac{\log R_n(x)}{n} < H+ \epsilon  ~~~~\forall ~~n \geq N_o(\epsilon,F_1), ~x~\in~F_1.$$
Hence we have,
$$\frac{\log R_n(x)}{H+\epsilon} < n
                      \Rightarrow R_n(x) < 2^{n(H+\epsilon)}.$$
Now, if $n_w$ is the window size and $n_w = 2^{\mathcal{L}_o(H+\epsilon)}$, then we have $R_{\mathcal{L}_o}(x) < n_w$ which implies that a \emph{match} of length $\mathcal{L}_o$ is bound to be found in $n_w$ symbols, hence the match length $$\mathcal{L}_o =\lfloor \frac{\log n_w}{H+\epsilon}\rfloor$$ is used in establishing the proof of optimality of SWLZ algorithm by Wyner and Ziv~\cite{Ziv94} and Jacob and Bansal~\cite{Bansal}. }
\end{CO}
\begin{CO}\label{CO2}
{Consider a zero entropy stationary and ergodic process s.t.
\begin{equation}\label{101}
\limsup_{n \rightarrow \infty} \frac{\log R_n(x)}{f(n)} = c.
\end{equation}
where $f(n) = o(n)$ is an invertible function of $n$ and $c$ is a constant. Here, $o(.)$ represents the small $o$ notation.

Now using Theorem \ref{OW2} by choosing $g(n) = f(n)$ we get the following corollary
\begin{cor}\label{splOWLT}
{If a stationary and ergodic process satisfies Eq. (\ref{101}) then we have with probability 1
$$ \lim_{n \rightarrow \infty} \frac{\log n}{f(L_n(x))} = c. $$
}
\end{cor}
Note that corollary \ref{OWLT} is a special case of corollary \ref{splOWLT} with $f(n) = \log n.$
By Eq. (\ref{101}) and Egoroff's Theorem~\cite{Royden68} in a similar manner as in Critical observation \ref{CO1}, we have for a given $\delta > 0$ and $\epsilon > 0$, $\exists$ a set $F_2$ of sequences $x$ with $\mu(F_2) > 1- \delta$ and an integer $N_o(\epsilon, F_2)$ s.t.
$$ 0 < \frac{\log R_n(x)}{f(n)} < c+ \epsilon  ~~~~\forall ~~n \geq N_o(\epsilon, F_2), ~x~\in~F_2.$$
Hence we have,
$$\frac{\log R_n(x)}{c+\epsilon} < f(n)
          \Rightarrow R_n(x) < 2^{(c+\epsilon)f(n)}.$$
Now, if $n_w$ is the window size and $n_w = 2^{(c+\epsilon)f(\mathcal{L}_o)}$, then we have $R_{\mathcal{L}_o}(x) < n_w$ which implies that a $\emph{match}$ of length $\mathcal{L}_o$ is bound to be found in $n_w$ symbols and hence the match length is given by
\begin{equation}\label{102}
\mathcal{L}_o = \lfloor f^{-1} (\frac{\log n_w}{c+\epsilon})\rfloor.
\end{equation}
for e.g. in the case of irrational rotation where as given by Eq. (\ref{2}) in Theorem \ref{KP}, $f(n) = \log n$ and $ c = 1$, we have
\begin{equation}\label{7}
\mathcal{L}_o = \lfloor{n_w}^{\frac{1}{1+\epsilon}}\rfloor.
\end{equation}
}
\end{CO}
\section{Convergence Rate of the Compression Ratio}\label{c34}
In ~\cite{Montano06}  the upper bounds and lower bounds on the compression ratio of the SWLZ algotrithm for Markov Sources with positive entropy have been obtained.  The upper bound is given by $H + O(\sqrt{\frac{\log{\log{n_w}}}{\log n_w}})$  and the lower bound is given by $ H + (H+o(1))\frac{\log \log n_w}{\log n_w}$.
Following the discussion in Section \ref{c33}, in this section we prove that a compression ratio upper bounded by $\frac{\log n_w}{{n_w}^a}$ , where $ a$ is a function of $n_w$ and approaches 1 as $n_w \rightarrow \infty$, for Fixed Shift Variant of SWLZ algorithm (FSLZ)~\cite{Bansal} and SWLZ algorithm is achievable for stationary and ergodic zero entropy processes generated by irrational rotation. Further, we give a general expression of the compression ratio for zero entropy cases under the setting described by Eq. (\ref{101}) for both FSLZ and SWLZ algorithms.

In the remainder of this section, we illustrate through the example of stationary and ergodic processes generated by an irrational rotation that a compression ratio of $\frac{\log n_w}{{n_w}^a}$ where $ a$ is a function of $n_w$ and approaches 1 as $n_w \rightarrow \infty$, is achieved for FDFS-LZ, FSLZ and SWLZ algorithms respectively following the proofs of optimality of these algorithms as given in~\cite{Ziv94} and \cite{Bansal} but choosing $\mathcal{L}_o$ given by Eq. (\ref{7}). It is emphasized here that the $\mathcal{L}_o$ chosen is in contrast as to what is chosen in \cite{Ziv94} and \cite{Bansal}. Moreover, following this illustration, an expression for the compression ratio for a class of zero entropy processes is given under a certain restriction on the convergence rate of the law given in Eq. (\ref{101}).

\subsection{{Fixed Database Fixed Shift Lempel-Ziv (FDFS-LZ) Algorithm}}\label{c34FDFS-LZ}
Consider a sequence $x = \{x_{n}\}_1^\infty$ which is sequentially made available to the encoder. Let $\mathcal{A}$ be the finite set of alphabet for the sequence $x$. Let $\beta \triangleq \lceil\log|\mathcal{A}|\rceil$. If $S$ is a finite set then $|S|$ denotes the cardinality of S. Consider a slight variant of Fixed Database Lempel Ziv (FDLZ) algorithm. Let us consider the string $x_1^N$ consisting of the first $N$ symbols of the sequence and a fixed database $y_1^{n_w}$ generated independently from the same source. The algorithm is described as follows :
\begin{enumerate}
  \item \emph{Partitioning: } Partition the sequence $x_{1}^{N}$ into $\mathcal{L}_o$ length blocks, where $\mathcal{L}_o$ is given by Eq. (\ref{7}). Hence, $\lfloor \frac{N}{\mathcal{L}_o} \rfloor$ blocks are created with $x_{\mathcal{L}_o\lfloor \frac{N}{\mathcal{L}_o} \rfloor + 1}^{N}$ as left over.
  \item \emph{Matching and Coding: } For $i^{th}$ block, (where $i = 1,2,.....\lfloor \frac{N}{\mathcal{L}_o} \rfloor$) try to find a match in the fixed database $y_1^{n_w}$. If a match is found, specify the pointer to its position in the fixed database. This requires $\lceil \log n_w \rceil$ bits and if no match is found, the block is not encoded. Use a 1 bit flag to specify if the block finds a match in the database. The remaining part $x_{\mathcal{L}_o\lfloor \frac{N}{\mathcal{L}_o} \rfloor + 1}^{N}$ is not encoded.

\end{enumerate}
Note that in this algorithm in contrast to FDLZ algorithm, we do not require the length of the block that finds a match to be encoded since it is already known to be $\mathcal{L}_o$. However, knowing $\mathcal{L}_o$ makes the algorithm universal only on a restricted class of stationary and ergodic processes.
The compression ratio for the algorithm described above is given by
\begin{equation}\label{R_FDLZ}
R_{FDFS-LZ} = \frac{1}{N}(m_1 (\lceil \log n_w \rceil + 1) + m_2(\beta \mathcal{L}_o+1) + \beta(N- \mathcal{L}_o\lfloor \frac{N}{\mathcal{L}_o} \rfloor)).
\end{equation}
where $m_1$ represents the number of blocks that find a match and $m_2$ represents those that do not find a match.

Now as an illustration we consider the example of irrational rotation to show that compression ratio for FDFS-LZ algorithm is upper bounded by $\frac{\log n_w}{{n_w}^a} + O(\frac{\log n_w}{n_w})$, where $a = 
\frac {1}{1+\frac {\log n_w}{n_w}}$.

Consider $x$, a sequence of $0's$ and $1's$ (produced through an irrational rotation) as described in section \ref{c33}. Let us fix a database of size $n_w$ . We now partition the sequence $x_1^{N}$ into segments of length $\mathcal{L}_o$ given by Eq. (\ref{7}). Now, by Egoroff's theorem, for a set $B$ with $\mu(B) > 1-\epsilon_{n_w}$ and for $x ~\epsilon~ B$, a match length of length $\mathcal{L}_o$ given by Eq. (\ref{7}) is guaranteed.

By total ergodicity of irrational rotation transformation, the frequency of $x, T'x, {T'}^{2}x,....$ belonging to $B$ is $\mu(B) > 1-\epsilon_{n_w}.$ for large $N$, where $T' = T^{\mathcal{L}_o}.$

So $1-\epsilon_{n_w}$ fraction of these phrases are guaranteed to be seen in the initial window. Without loss of generality, assuming $\frac{N}{\mathcal{L}_o}$ to be an integer, the compression ratio is given by
\begin{equation}\label{105}
\begin{split}
R_{FDFS-LZ} &< \frac{\frac{N}{\mathcal{L}_o}(1-\epsilon_{n_w})(\lceil \log n_w \rceil + 1) + \epsilon_{n_w}(\beta N +\frac{N}{\mathcal{L}_o})}{N}\\
         &< \frac{(1-\epsilon_{n_w})(\log n_w + 2)}{{n_w}^{\frac{1}{1+\epsilon_{n_w}}}-1} +\epsilon_{n_w} (\beta+ \frac{1}{\mathcal{L}_o}).\\ as ~N \rightarrow \infty .
\end{split}
\end{equation}
Now we can always choose $\epsilon_{n_w} = \frac{\log n_w}{n_w}.$ Therefore, for this choice of $\epsilon_{n_w}$, we have the following theorem,

\begin{mythm}\label{res1}
{For the above choice of $\epsilon_{n_w}$ we can choose $N$ large enough, so that the compression ratio of the FDFS-LZ algorithm with fixed match lengths is upper bounded by $\frac{\log n_w}{{n_w}^a} + O(\frac{\log n_w}{n_w}) $, where $a = \frac{1}{1+\frac{\log n_w}{n_w}}$. 
}
\end{mythm}
\subsection{{Fixed Shift Variant of Lempel-Ziv (FSLZ) Algorithm}}\label{c34FSLZ}
We first consider FSLZ algorithm introduced in~\cite{Bansal} to gain insight into the working of the universal SWLZ algorithm. Let us consider the string $x_1^N$ consisting of the first $N$ symbols of the sequence. Following the discussion in Section \ref{c33} and using critical observation \ref{CO2}, let us take for $\epsilon > 0$, the match length given by Eq. (\ref{7}).
Note the contrast in the choice of match length that is used in~\cite{Ziv94} and~\cite{Bansal} for positive entropy stationary and ergodic processes.The algorithm is described as follows:
\begin{enumerate}
  \item  \emph{Initialization}: A window of size $n_w$ is fixed. Let $j =1$ and $n_j = n_w$. The first $n_w$ symbols are transmitted without compression. Let $x_1^{n_1}$ be the current window.
  \item  \emph{Matching}: At the $j$th step if $n_j+\mathcal{L}_o > N$, terminate the algorithm, else if $x_{n_j+1}^{n_j+\mathcal{L}_o} = x_{n_j-n_w+k}^{n_j-n_w+k+\mathcal{L}_o-1}$ for any $k ~\in ~{1,2,....,n_w}$ then there is a match, i.e., sequence $x_{n_j+1}^{n_j+\mathcal{L}_o}$ has a match in the current window.
  \item \emph{Coding}: If there is a match, let $s_j$ be the location of the match in the current window. Then $\lceil \log n_w\rceil$ bits are required to code $s_j$. If a match is not found the number of bits required is $\beta \mathcal{L}_o$ ( $\beta \triangleq \lceil \log|\mathcal{A}|\rceil, |\mathcal{A}|$ is the alphabet size). In addition, a one bit flag is used to specify if the match is found or not found.
  \item \emph{Sliding}: $n_{j+1}= n_{j}+\mathcal{L}_o$, i.e., the current window now becomes $x_{n_{j+1}-n_w+1}^{n_{j+1}}$.
\end{enumerate}
Repeat the steps 2, 3 and 4 until the sequence $x_1^N$ terminates at the $(m+1)$-th step. The remaining $N-n_w-m\mathcal{L}_o <\mathcal{L}_o$ symbols are encoded without compression. Thus, a total of $N-m\mathcal{L}_o < n_w+\mathcal{L}_o$ are encoded without any compression. Let $m_1$ be the number of blocks that have a match and $m_2$ be the number of blocks that do not have a match. One bit per block is used to denote if it is good or bad. Hence, taking overheads into account, the number of bits per symbol required by the code is given by
\begin{equation}\label{8}
\begin{split}
R_{FSLZ} &= \frac{1}{N}\Big[(N - m\mathcal{L}_o)\beta + m_1\lceil \log n_w \rceil + m_2\beta \mathcal{L}_o +m\Big]\\&
< \frac{1}{N}\Big[(n_w + \mathcal{L}_o)\beta + m_1\lceil \log n_w \rceil + m_2\beta \mathcal{L}_o + m\Big].
\end{split}
\end{equation}
The first term in the expression converges to 0 as $N \rightarrow \infty$.
Also, we have
\begin{equation}\label{9}
m_1 \leq m < \frac{N}{\mathcal{L}_o}.
\end{equation}
Using Eq. (\ref{7}) and (\ref{9}) we have,
\begin{equation}\label{10}
\frac{m_1\lceil \log n_w\rceil}{N} < \frac{\lceil \log n_w\rceil}{\mathcal{L}_o} < \frac{\lceil \log n_w\rceil}{{n_w}^{\frac{1}{1+\epsilon}}-1}.
\end{equation}
which converges to 0 as $n_w \rightarrow \infty$.
Therefore, using Eq. (\ref{10}) we have
\begin{equation}\label{11}
  \lim_{n_w \rightarrow \infty} \frac{m_1\lceil \log n_w\rceil}{N} = 0.
\end{equation}
 Hence, we have the following
\begin{CO}\label{CO3}
{ The second term in the expression of $R_{FSLZ}$ (Eq. (\ref{8})) converges to $0$ for all $\epsilon > 0 $ as $n_w\rightarrow\infty$ and the convergence rate is given by $O(\frac{\log n_w}{n_w^{\frac{1}{1+\epsilon}}})$.
}
\end{CO}
Now , define $ G = \{ x: R_{\mathcal{L}_o}(x) > n_w\}.$ Thus, if $ x~\in~ G$, then $x$ does not have a match in the previous $n_w$ symbols. We use $1_G$ to denote the indicator function of $G$. Since $T$ is \emph{totally} ergodic (i.e., $T^k$ is ergodic for all k as $k\theta$ is irrational for all $k$), Z the \emph{shift transformation} is also totally ergodic. By the Ergodic Theorem~\cite{Shields96} we have (almost surely)
\begin{align}\label{12}
\frac{m_2\mathcal{L}_o}{N} < \frac{m_2}{m}
                 &= \frac{\sum_{j=0}^{m-1} 1_G(Z^{j\mathcal{L}_o+n_w}(x))}{m}
                  \rightarrow \mu(G).
\end{align}
Now,
\begin{equation*}
\begin{split}
G &= \{ x: R_{\mathcal{L}_o}(x) > n_w \} \\
  &\subset \{ x: \frac{\log R_{\mathcal{L}_o}(x)}{\log \mathcal{L}_o} > 1+\epsilon \}. \\
\end{split}
\end{equation*}
Hence, by Theorem \ref{KP} $\lim_{n_w \rightarrow \infty} \mu(G) = 0$.
\begin{CO}\label{CO4}
{It is established in~\cite[Sec. 2 page-3944 (Proof of Theorem 1.1), Sec. 3 page-3948, 3949 (Proof of Theorem 1.3)]{Kim} that $~\forall ~\epsilon ~ >~ 0$, the convergence rate of the probability of the set $G$ is at least as fast as $\frac{1}{{n_w}^{b(\epsilon)}}$  $\forall ~n_w \geq n_w^o(\epsilon)$, where $b(\epsilon) > \frac{1}{1+\epsilon}$. Now for a given $n_w$, we choose a \emph{minimum} $\epsilon$ such that $n_w \geq n_w^o(\epsilon) $. We denote this $\epsilon$ by $\epsilon_{n_w}$. (Note that as $n_w \rightarrow \infty, \epsilon_{n_w} \rightarrow 0$) 
}
\end{CO}

Combining (\ref{8}), (\ref{9}), (\ref{11}), (\ref{12}) and $\mu(G) \rightarrow 0$, we have for almost every sequence generated by an irrational rotation, the FSLZ algorithm is asymptotically optimal. i.e.,
$\lim_{n_w \rightarrow \infty} \lim_{N\rightarrow \infty} R_{FSLZ} = 0$.

Using critical observations \ref{CO3} and \ref{CO4}, we can now state the following theorem,\\
\begin{mythm}\label{res2}
{For a stationary and ergodic process generated by an irrational rotation a compression ratio given by $O(\frac{\log n_w}{{n_w}^a})$ is achieved by FSLZ algorithm, where $a = \frac{1}{1+\epsilon_{n_w}}$. 
}
\end{mythm}

\subsection{{Sliding Window Lempel-Ziv (SWLZ) Algorithm}}\label{c34SWLZ}
Sliding Window LZ is described as follows~\cite{Ziv94}\cite{Bansal} :
\begin{enumerate}
  \item  \emph{Initialization}: A window of size $n_w$ is fixed. Let $j =1$ and $n_j = n_w$. The first $n_w$ symbols are transmitted without compression. Let $x_1^{n_1}$ be the current window.
  \item  \emph{Matching}: At the $j$th step let $L_j$ be the largest integer such that the copy of $x_{n_j+1}^{n_w+L_j}$ begins in the current window and $n_j+L_j\leq N$. Let $s_j$ denote the starting index of the match in the current window. The matched phrase $x_{n_j+1}^{n_w+L_j}$ is denoted by $x^{(j)}$. If a match is not found $ L_j =1$, $s_j = 0$.
  \item \emph{Coding}: If $s_j >0$, the length $L_j$ of the matched code can be specified using $\gamma\log (L_j+1)$ bits using the integer code described in~\cite{Ziv94}, here $\gamma$ is some constant. The matched location $s_j$ can be specified using $\lceil\log n_w\rceil$ bits. If $s_j =0$ or if, using the above procedure, the total number of bits needed to represent a phrase exceeds $\beta L_j$, then $\beta L_j$ bits are used for encoding. A one bit flag is used to denote which of these two encoding schemes was used.
  \item \emph{Sliding}: For the next window $n_{j+1} = n_j + L_j$ and the window for the next iteration is $x_{n_{j+1}-n_w+1}^{n_{j+1}}$.

\end{enumerate}
Repeat the steps 2, 3 and 4 until the sequence $x_1^N$ is exhausted.
Let $B(x^{(j)}) + 1$ denote the number of bits required to encode the $j$-th phrase. Then, we have
\begin{equation}\label{5}
    B(x^{(j)}) = \min{\{\gamma\log (L_j+1) + \lceil\log n_w\rceil, \beta L_j \}}.
\end{equation}
If the total number of phrases is $c(N)$ then, the number of bits required per symbol or the compression ratio is:
\begin{equation}\label{6}
  R_{SWLZ}= \frac{1}{N}(n_w\beta +\sum_{j=1}^{c(N)} B(x^{(j)})+c(N)).
\end{equation}
Now we analyze the performance of SWLZ algorithm on zero entropy sequences generated by an irrational rotation. The idea behind the analysis is inspired by the method of partitioning of sequences used in~\cite{Ziv94}. However, unlike them the class of partitions considered here are obtained by shifting the partition used by them. By this the optimality of SWLZ algorithm is obtained in almost sure sense in~\cite{Bansal}. Consider the intervals defined as $ I_r = [r+n_w,r+n_w+\mathcal{L}_o-1],~ r = 1,2,...,N' $. Here, $ N' = N-\mathcal{L}_o-n_w +1.$ Interval $I_r$ is bad if a copy of ${(x_i)}_{i ~\in~ I_r}$ does not begin in the string $ n_w -\mathcal{L}_o $ symbols preceeding it. Let $m$ be the number of such bad intervals.
Now, we define
$ G = \{x:R_{\mathcal{L}_o}(x) > n_w - \mathcal{L}_o\}.$
So , if $ x ~\in~ G $, then $[1,\mathcal{L}_o]$ is a bad interval.
By Ergodic Theorem almost surely,
\begin{equation}\label{15}
\frac{m}{N} < \frac{m}{N'} = \frac{\sum_{j=0}^{N'-1}1_G(Z^j(x))}{N'} \rightarrow \mu (G). ~~ (as ~N \rightarrow \infty)
\end{equation}
Now, $\quad \quad ~~~G = \{ x: R_{\mathcal{L}_o}(x) > n_w - \mathcal{L}_o \}$\\
\begin{equation*}
 \begin{split}
  &\subset \{ x: R_{\mathcal{L}_o}(x) > \mathcal{L}_o^{1+\frac{\epsilon}{2}}  \}\\& = \{ x: \frac{\log R_{\mathcal{L}_o}(x)}{\log \mathcal{L}_o} > 1+ \frac{\epsilon}{2} \}.
\end{split}
\end{equation*}
Hence by Theorem \ref{KP},
\begin{equation}\label{18}
\lim_{n_w \rightarrow \infty} \mu(G) = 0.
\end{equation}
which implies from (\ref{15})
$\lim_{n_w \rightarrow \infty} \frac{m}{N} = 0.$
As, already specified in critical observation \ref{CO4}, we have the convergence of rate of $\mu(G)$ to $0$ (as $n_w \rightarrow \infty$) is at least as fast as $\frac{1}{{n_w}^b},$ where $b > a$.

Let us consider the SWLZ parsing of the sequence $x_1^N$. A phrase is defined to be internal with respect to interval $I_r$ if it begins and ends in the interval $[r+n_w, r+ \mathcal{L}_o+n_w-2]$~\cite{Ziv94}. Moreover, an interval $I_r$ with an internal phrase is bad~\cite{Ziv94}.

We define for $k = 0,1,2.....,\mathcal{L}_o - 1$ ,
$ p_k = \lfloor \frac{ N- n_w - k}{\mathcal{L}_o} \rfloor. $ and let $ \mathcal{P}_k = \{ I_{k + 1}, I_{k + \mathcal{L}_o + 1},....,I_{k + (p_k-1)\mathcal{L}_o + 1}\}.$ $\mathcal{P}_k$ covers the interval $[n_w + 1, N]$ except for an initial segment $[n_w + 1, n_w + k]$ and final segment $[N_k, N]$ where $ N_k = n_w + k + p_k\mathcal{L}_o + 1 $, each of these uncovered segments are of length less than $\mathcal{L}_o$, since $ k = 0, 1, 2,..., \mathcal{L}_o - 1.$

Let the collection of internal phrases corresponding to $\mathcal{P}_k$ together with initial and final segment be denoted by set $S_k$ and let $m_k$ be the number of bad intervals. Then, we have
\begin{equation}\label{19}
\sum_{j ~\in~ S_k} L_j \leq \mathcal{L}_o(m_k + 2).
\end{equation}
Since an internal phrase can only be supported on a bad interval, let us define~
$m_{k_o} = \min_k \{m_k\},$
 where $m = \sum_{k=0}^{\mathcal{L}_o-1} m_k$.
Hence, we have using the definition of $m_{k_o}$ and the above expression for $m$.
\begin{equation}\label{22}
\mathcal{L}_om_{k_o} \leq m.
\end{equation}
Using Eq. (\ref{6}), the number of bits per symbol for SWLZ algorithm is given by
\begin{equation}\label{23}
R_{SWLZ}= \frac{1}{N} (n_w\beta +\sum_{j ~\in~ S_{k_o}} (\beta L_j+1) + \sum_{j ~\in~ S_{k_o}^c} (B(x^{(j)})+1)).
\end{equation}
Using Eq. (\ref{19}) and (\ref{22}) we have
\begin{equation}\label{24}
\begin{split}
R_{SWLZ} \leq \frac{1}{N} ((n_w + 2 \mathcal{L}_o)(\beta + 1) +(\beta+1) m \\
  +\sum_{j ~\in~S_{k_o}^c} B(x^{(j)})+|{S_{k_o}}^c|).
\end{split}
\end{equation}
 The terms $\frac{n_w(\beta + 1)}{N}$ and $\frac{2(\beta+1) \mathcal{L}_o}{N}$ converge to 0 as $N \rightarrow \infty$. From Eq. (\ref{18}) it is evident that the second term in the expression for $R_{SWLZ}$ given by Eq. (\ref{24}) converges to 0 as ($ N \rightarrow \infty $ followed by $n_w \rightarrow \infty$). Now, we consider the last term in the expression for $R_{SWLZ}$ given in Eq. (\ref{24}) i.e., ~$\frac{1}{N} \sum_{j ~\in~ S_{k_o}^c} B(x^{(j)}).$
Using the method used in~\cite{Ziv94}, let $ d =|S_{k_o}^c| $.
Since, $S_{k_o}^c$ comprises of non internal phrases corresponding to partition $\mathcal{P}_{k_o}$ together with the initial and final segments,the non-internal phrase should end at the last index of any interval that belongs to $\mathcal{P}_{k_o}$.
\begin{equation}\label{27}
|S_{k_o}^c| = d \leq \frac {N'}{\mathcal{L}_o} < \frac{N}{\mathcal{L}_o}.
\end{equation}
Using Eq. (\ref{27}) we have $\lim_{n_w \rightarrow \infty} \frac{|{S_{k_o}}^c|}{N} = 0 $. Further, we have from~\cite{Ziv94}
\begin{equation}\label{28}
\begin{split}
\frac{1}{N} \sum_{j \epsilon S_{k_o}^c} B(x^{(j)}) &= \frac{1}{N}\sum_{j ~\in~ S_{k_o}^c}\Big\{ \gamma \log(L_j + 1) + \lceil \log n_w \rceil\Big\}  \\
                                      &= \frac{1}{N}|S_{k_o}^c| \lceil \log n_w \rceil  + \frac{1}{N}\sum_{j ~\in~ S_{k_o}^c} \gamma \log(L_j + 1)  \\
                                      &\leq \frac{1}{\mathcal{L}_o} \lceil \log n_w \rceil + \frac{d\gamma}{N} \sum_{j~ \in~ S_{k_o}^c} \frac{1}{d}\log(L_j + 1) ~(a_1)\\
                                      &\leq \frac{\lceil \log n_w \rceil}{\mathcal{L}_o} + \frac{d\gamma}{N} \log(\frac{1}{d} \sum_{j ~\in~ S_{k_o}^c} (L_j + 1))  ~~(a_2)\\
                                       &\leq \frac{\lceil \log n_w \rceil}{n_w^{\frac{1}{1+\epsilon}}-1} + \frac{\gamma}{\mathcal{L}_o} \log(\mathcal{L}_o + 1) ~~(a_1)\\
                                      &= 0. ~~(as ~n_w \rightarrow \infty)
\end{split}
\end{equation}
Here, ($a_1$) means using Eq. (\ref{27}) and ($a_2$) means using the concavity of log function.
\begin{CO}\label{CO5}
{It is evident in Eq. (\ref{28}) that the compression ratio converges to 0 for good phrases at the rate given by $O(\frac{\log n_w}{n_w^{\frac{1}{1+\epsilon}}})$, for every $\epsilon > 0.$
}
\end{CO}
Hence, combining results of Eq. (\ref{18}) and (\ref{28}) we get for almost every sequence generated by an irrational rotation, the SWLZ algorithm is asymptotically optimal i.e.,
$\lim_{n_w \rightarrow \infty} \lim_{N \rightarrow \infty} R_{SWLZ}(x) = 0 ~a.s.$\\
So using critical observations \ref{CO4} and \ref{CO5} we have the following theorem,\\
\begin{mythm}\label{res4}
{For stationary and ergodic processes generated by an irrational rotation a compression ratio given by $O(\frac{\log n_w}{{n_w}^a})$ is achieved by SWLZ algorithm, where $a = \frac{1}{1+\epsilon_{n_w}}$.
}
\end{mythm}
\subsection{Possible Extensions}
One can consider a general setting where the process is assumed to be stationary and totally ergodic with zero entropy and the asymptotics of the first return time follow the law given by Eq. (\ref{101}). Here, we can choose $\mathcal{L}_o$ given by Eq. (\ref{102}). In such a case following the proofs considered above in subsections 4.2 and 4.3, we can get convergence rates given by $ O(\frac{\log n_w}{f^{-1}(\frac{\log n_w}{c+\epsilon_{n_w}})})$ (Note: $f(n) \geq \log n$ always holds true \cite{Kim11}). For instance, if $f(n) =\sqrt{n}$ the compression ratio for FSLZ algorithm is given by $O(\frac{{(c+\epsilon_{n_w})}^2}{\log n_w}).$

In the subsequent part of the paper, we discsuss about the large deviation property for recurrence times and match lengths. Moreover, we also discuss about entropy estimation using recurrence times. Before establishing these results we need some preliminary results on statsitics of recurrence times which we present in the next section. 

Further, in this paper unless stated otherwise $i)$ we use $H$ to represent the entropy rate of the source $X$. $ii)$ A $\phi$-mixing process is assumed to be $\phi$-mixing in both forward and backward directions.
\section{{Recurrence Time Statistics }}\label{c43}
\begin{mythm}\label{KimT}
{\textit{(Kim's Theorem \cite{Kim12})}
For a $X$ satisfying $\psi$-mixing or $\phi$-mixing condition (for definiton of $\psi$ and $\phi$- mixing conditons refer appendix) with summable coefficients,
\begin{equation}\label{LB}
\begin{split}
\mu(R_n(X) > t|X_1^n = x_1^n) > \xi_x e^{-t\xi_x\mu(x_1^n)}(1-\\2\sqrt{C_x(\xi_x\mu(x_1^n)t\vee 1)})
~~~\forall~~ t > 0 .
\end{split}
\end{equation}
\begin{equation}\label{UB}
\begin{split}
\mu(R_n(X) > t|X_1^n = x_1^n) &< \xi_x e^{-t\xi_x\mu(x_1^n)}[1+ K(x,t)+\\&+ 2C_x(\xi_xt\mu(x_1^n)\vee 1)]~\forall~ t \geq \rho_x.
\end{split}
\end{equation}
}
\end{mythm}
where $ C_x = C\{\inf_{n\leq \triangle\leq 1/\mu(x_1^n)}[\triangle \mu(x_1^n)+ \ast(\triangle)]\}$ ($C>0$ is a constant, $\ast$ represents $\psi$ or $\phi$), $\rho_x = \frac{2\sqrt{C_x}}{(\sqrt{1+C_x}+\sqrt{C_x})\xi_x\mu(x_1^n)}$.$$ K(x,t) = 2\sqrt{C_x(t\xi_x\mu(x_1^n)\vee 1)(1+C_x(t\xi_x\mu(x_1^n)\vee 1))},$$   and $C_x \rightarrow 0 ~ (as ~ n \rightarrow \infty)$ and $\xi_x~\in~[E_1, E_2], (0 < E_1<1<E_2<\infty).$ $a_1\vee a_2$ means $\max\{a_1,a_2\}$. Following additional properties as listed in \cite{Kim12} and originally proved in \cite{Collet99}\cite{Abadi01} hold for $\phi$-mixing processes:
\begin{enumerate}
\item For an exponentially $\phi$-mixing process,~ $\forall~ x_1^n ~\in~ \mathcal{A}^n,$ there exists a positive constant $D_o$ and $\Gamma > 0$, s.t. $\forall~ n \geq n_o$
\begin{equation}\label{exp_phi}
C_x \leq D_oe^{-\Gamma n}.
\end{equation}
\item Let $B_n(s)$ be the set of $x_1^n ~\in~\mathcal{A}^{n}$; such that $R_n(x) < \frac{n}{s}.$ Then, for any $\phi$-mixing process, there exists $s~\in~\mathcal{N}$ ($\mathcal{N}$ being the set of natural numbers), and two positive constants $D_1$ and $d_1$ such that
\begin{equation}\label{4}
\mu(\{x: x_1^n ~\in~B_n(s)\}) \leq D_1e^{-d_1n}.
\end{equation}
\item For exponentially $\phi$-mixing processes for every $x_1^n ~ \in~ {\mathcal{A}}^n\backslash B_n(s)$
\begin{equation}\label{xi}
|\xi_x - 1| < D_2e^{-d_2n}.~~~~(for~n~large~enough)
\end{equation}
Here, $D_2$ and $d_2$ are constants.
\end{enumerate}
Now, we state a Lemma which is required in the proof of Theorem \ref{LDP2} stated in section \ref{c44}.
Let $A_1$, $A_2$ and $A_3$ be three sets such that $A_3 = A_1\cap A_2$. Suppose $\mu(A_1) > 1- p_1e^{-p_2n}$
and $\mu(A_2) > 1-q_1e^{-q_2n}$, where $p_1$, $p_2$, $q_1$ and $q_2$ are positive constants. Then, we have
\begin{lem}\label{L1}
{ $\mu(A_3) > 1- (p_1 + q_1)e^{-\min\{p_2,q_2\}n}$.}
\end{lem}
Lemma \ref{L1} is proved in the appendix.\\
{\it \bf Definition} \cite{Shields96} \\
A stationary and ergodic process is said to have exponential rates for entropy if for every $\epsilon~ > ~0$, we have
\begin{equation}\label{LDP_Sh}
\mu(\{x_1^n: 2^{-n(H+\epsilon)} \leq \mu(x_1^n) \leq 2^{-n(H-\epsilon)}\}) \geq 1- r(\epsilon,n).
\end{equation}
where $-\frac{1}{n}\ln r(\epsilon, n)$ is bounded away from 0 or in other words $r(\epsilon, n) = e^{-k(\epsilon)n}$, where $k(\epsilon)$ is a real valued positive function of $\epsilon$.

In the next section, we state our main results on LDP of recurrence times. 
\section{{Main Theorems on LDP of recurrence times}}\label{c44}
\begin{mythm}\label{LDP1}{ For a process satisfying $\psi$-mixing condition or $\phi$-mixing condition with summable coefficients and with exponential rates for entropy,
$$\mu(\frac{\log R_n(X)}{n} > H+ \epsilon) \leq e^{-f(\epsilon)n} ~~~~  \forall~ n \geq N(\epsilon).$$}\end{mythm}
where, $f(\epsilon)$ is a real positive valued function for all $\epsilon~ > ~0 $ and $f(0) = 0$.
\begin{cor}\label{CO42}{
Under the conditions of Theorem \ref{LDP1}, we have
$$\mu(\frac{\log m}{L_m(X)} > H+ \epsilon) \leq e^{-f(\epsilon)\frac{\log m}{H+\epsilon}} ~~~~  \forall~ m \geq M(\epsilon).$$}\end{cor}
\begin{mythm}\label{LDP2}
{ For an exponentially $\phi$-mixing process,
$$\mu(\frac{\log R_n(X)}{n} < H-\epsilon) \leq e^{-g(\epsilon)n} ~~~~\forall~ n \geq N'(\epsilon).$$}\end{mythm}
where $g(\epsilon)$ is a real positive valued function for all $\epsilon~ > ~0 $ and $g(0) = 0$.
\begin{cor}\label{CO43}
{Under the conditions of Theorem \ref{LDP2}, we have
$$\mu(\frac{\log m}{L_m(X)} < H- \epsilon) \leq e^{-g(\epsilon)\frac{\log m}{H-\epsilon}} ~~~~  \forall~ m \geq M'(\epsilon).$$
}\end{cor}
Theorem \ref{LDP1} and \ref{LDP2} are combined in the form of
\begin{mythm}\label{LDP3}
{\textit{(Large Deviation Property for Recurrence Times)}
For an exponentially $\phi$-mixing process with exponential rates for entropy,
$$ \mu(|\frac{\log R_n(X)}{n} - H| > \epsilon) \leq 2e^{-I(\epsilon)n} ~~~~\forall n~ \geq N''(\epsilon). $$
}\end{mythm}
where, $I(\epsilon)= \min\{f(\epsilon), g(\epsilon)\}$ and $N''(\epsilon) =  \max\{N(\epsilon), N'(\epsilon)\}$.\\
\begin{Rem}\label{rem2}
{An aperiodic and irrecucible Markov Chain is exponentially $\phi$-mixing and has exponential rates for entropy \cite{Bradley}\cite{Shields96}. Therefore, using Theorem \ref{LDP3}, it can be inferred that the quantity $\frac{\log R_n(X)}{n}$ for an aperiodic and irreducible Markov chain satisfies Large Deviation Property.
}\end{Rem}
\section{{Proofs of LDP Theorems on Recurrence Times}}\label{c45}
\textit{Proof of Theorem \ref{LDP1}:}
Let $A_n^{(\delta)}$ be a set of $n$ long sequences defined as,
$$ A_n^{(\delta)} = \{x_1^n: 2^{-n(H+\delta)} \leq \mu(x_1^n) \leq 2^{-n(H-\delta)}\}.$$Now,
\begin{equation}
\begin{split}
&\mu(\frac{\log R_n(X)}{n} > H+ \epsilon) = \mu(R_n(X) > 2^{n(H+\epsilon)}) \\&= \sum_{y~\in~ \mathcal{A}^n}\mu(y)\mu(R_n(X) > 2^{n(H+\epsilon)}|X_1^n = y)\\
&= \sum_{y~\in~ A_n^{(\delta)}}\mu(y)\mu(R_n(X) > 2^{n(H+\epsilon)}|X_1^n = y)\\& + \sum_{y~\in~ {A_n^{(\delta)}}^c}\mu(y)\mu(R_n(X) > 2^{n(H+\epsilon)}|X_1^n = y)\\
&< \sum_{y~\in~ A_n^{(\delta)}}\mu(y)[\xi_y e^{-2^{n(H+\epsilon)}\xi_y\mu(y)}[1+ K(y,2^{n(H+\epsilon)})\\&+ 2C_y(\xi_y2^{n(H+\epsilon)}\mu(y)\vee 1)]] + \sum_{y~\in~ {A_n^{(\delta)}}^c}\mu(y)~~~~~ (a)
\end{split}
\end{equation}
\begin{equation}\label{main1}
\begin{split}
&< \sum_{y~\in~ {A_n}^{(\delta)}}\mu(y)[E_2 e^{-2^{n(H+\epsilon)}E_1\mu(y)}[1+  V\\&+  2d(E_22^{n(H+\epsilon)}\mu(y)\vee 1)]  +  \sum_{y~\in~ {A_n^{(\delta)}}^c}\mu(y)~~~~~~~~ (b)
\end{split}
\end{equation}
where
$$
V = 2\sqrt{d(2^{n(H+\epsilon)}E_2\mu(y)\vee 1)(1+d(2^{n(H+\epsilon)}E_2\mu(y)\vee 1))}.$$
$(a)$ follows from the use of inequality (\ref{UB}) and Remark \ref{rem6} as stated in Appendix. $(b)$ follows from using the fact that $\xi_y ~\in~[E_1,E_2]$ and $C_y \rightarrow 0 ~as~n \rightarrow \infty \Rightarrow C_y < d ~ \forall$ $y$ and $n$ large enough, where $d > 0$ is an arbitrary constant. For $y ~\in~ A_n^{(\delta)}$, we have $$2^{n(\epsilon-\delta)}\leq 2^{n(H+\epsilon)}\mu(y) \leq 2^{n(\epsilon + \delta)}.$$ For every $\epsilon > 0$, choose $\delta = \frac{\epsilon}{2}$. Consequently, we have
\begin{equation}\label{sub1}
2^{\frac{n\epsilon}{2}} \leq 2^{n(H+\epsilon)}\mu(y) \leq 2^{\frac{3n\epsilon}{2}} ~ \forall ~y ~\in~A_n^{(\frac{\epsilon}{2})}.
\end{equation}
Also, $2^{\frac{3n\epsilon}{2}}E_2 > 1$ since $E_2 > 1$. Hence, using (\ref{main1}) and (\ref{sub1}) we have,
\begin{equation}\label{final1}
\begin{split}
\mu(\frac{\log R_n(X)}{n} > H+ \epsilon) \leq \sum_{y~\in~ A_n^{(\frac{\epsilon}{2})}}\mu(y)[E_2e^{-E_12^{\frac{n\epsilon}{2}}}(1+\\2\sqrt{dE_22^{\frac{3n\epsilon}{2}}(1+dE_22^{\frac{3n\epsilon}{2}})}+ 2dE_22^{\frac{3n\epsilon}{2}})] +  \sum_{y~\in~ {A_n^{(\frac{\epsilon}{2})}}^c}\mu(y).
\end{split}
\end{equation}
Using (\ref{LDP_Sh}) and (\ref{final1}), for processes having exponential rates for entropy and satisfying $\psi$-mixing condition or $\phi$-mixing condition with summable coefficients, we have
\begin{equation}\label{final_r1}
\begin{split}
&\mu(\frac{\log R_n(X)}{n} > H+ \epsilon) \leq  \Big\{E_2e^{-E_12^{\frac{n\epsilon}{2}}}[1+ 2dE_22^{\frac{3n\epsilon}{2}}\\&+2\sqrt{dE_22^{\frac{3n\epsilon}{2}}(1+dE_22^{\frac{3n\epsilon}{2}})}]\Big\} + r(\frac{\epsilon}{2},n)
< e^{-f(\epsilon)n}.
\end{split}
\end{equation}
This completes the proof of Theorem \ref{LDP1}.\\
\begin{Rem}\label{rem3}
{ Since the first term on the right hand side of inequality (\ref{final_r1}) stated above rapidly (super exponentially) converges to $0$, $f(\epsilon)$ behaves in a similar manner as $-\frac{\ln r(\frac{\epsilon}{2},n)}{n} = k(\frac{\epsilon}{2})$ (Also see Remark \ref{rem7} as stated in appendix).
}\end{Rem}
{\it Proof of Corollary \ref{CO42}:}
From Observation \ref{Obs41}, we have $$R_n(x) > 2^{n(H+\epsilon)} \Leftrightarrow L_{2^{n(H+\epsilon)}}(x) < n $$
$$\Rightarrow \mu(L_{2^{n(H+\epsilon)}}(X) < n) = \mu(R_n(X) > 2^{n(H+\epsilon)}) < e^{-f(\epsilon)n}.$$
$\forall~n\geq N(\epsilon)$.
Now, letting $m = 2^{n(H+\epsilon)}$, we have
\begin{equation}
\begin{split}
\mu(L_{m}(X) < \frac{\log m}{H + \epsilon}) < e^{-f(\epsilon)\frac{\log m}{H + \epsilon}}   ~~~~\forall ~m \geq M(\epsilon)\\
\Rightarrow \mu(\frac{\log m}{L_m(X)} > H+ \epsilon) < e^{-f(\epsilon)\frac{\log m}{H + \epsilon}}.
\end{split}
\end{equation}
{\it Proof of Theorem \ref{LDP2}:}
Let $A_n^{(\frac{\epsilon}{2})}$ be the same set as considered in the proof of Theorem \ref{LDP1}. For each $y~\in~A_n^{(\frac{\epsilon}{2})}$, we have
\begin{equation}\label{sub2}
2^{-\frac{3n\epsilon}{2}} \leq \mu(y)2^{n(H-\epsilon)} \leq 2^{-\frac{n\epsilon}{2}}.
\end{equation}
Now,
\begin{equation}\label{final2}
\begin{split}
&\mu(\frac{\log R_n(X)}{n} < H-\epsilon)
= 1 - \mu(\frac{\log R_n(X)}{n} \geq H-\epsilon)\\
                                  &\leq 1-\mu(\frac{\log R_n(X)}{n} > H-\epsilon)\\
                                    &=1-\mu(R_n(X) > 2^{n(H-\epsilon)})\\
                                   &=1-\sum_{y~\in~ \mathcal{A}^n}\mu(y) \mu(R_n(X) > 2^{n(H-\epsilon)}|X_1^n = y)\\
                                      &< 1-\sum_{y~\in~\mathcal{A}^n}\mu(y)[\xi_ye^{-\mu(y)\xi_y2^{n(H-\epsilon)}}(1- \\& 2\sqrt{C_y(\xi_y\mu(y)2^{n(H-\epsilon)}\vee 1)})]~~(a) \\
                                      &< 1-\sum_{y~\in~ A_n^{(\frac{\epsilon}{2})}}\mu(y)[\xi_ye^{-E_22^{-\frac{n\epsilon}{2}}}(1- 2\sqrt{C_y(E_22^{-\frac{n\epsilon}{2}}\vee 1)})] ~(b) \\
                                      &= 1- \sum_{y~\in~ A_n^{(\frac{\epsilon}{2})}}\mu(y)[\xi_ye^{-E_22^{-\frac{n\epsilon}{2}}}(1-2\sqrt{C_y})]~(c)
\end{split}
\end{equation}
Here, $(a)$ follows from (\ref{LB}), $(b)$ follows from the fact that $\xi_y ~\in~[E_1,E_2]$ and inequality (\ref{sub2}). Also in $(b)$ the negative term contributed by sequences belonging to the set ${A_n^{(\frac{\epsilon}{2})}}^c$ is ignored because we are looking at an upper bound. $(c)$ follows because eventually $E_22^{-\frac{n\epsilon}{2}} < 1$, since $E_22^{-\frac{n\epsilon}{2}} \rightarrow 0~(as ~ n~\rightarrow ~\infty).$

To proceed further, we introduce the following notations, let $A_1 = \mathcal{A}^n \backslash B_n(s); A_2 = A_n^{(\frac{\epsilon}{2})}.$ From (\ref{4}) and (\ref{LDP_Sh}), we have $\mu(A_1) > 1- D_1e^{-d_1n}$ and $\mu(A_2) > 1- e^{-k(\frac{\epsilon}{2})n}$ respectively for processes with exponential rates for entropy. Let $A_3 = A_1 \cap A_2$.

Therefore from (\ref{final2}), we have
\begin{equation}\label{final_r2_sub}
\begin{split}
&\mu(\frac{\log R_n(X)}{n} < H -\epsilon) < 1-\\&\sum_{y~\in~ A_3}\mu(y)\xi_ye^{-E_22^{-\frac{n\epsilon}{2}}}(1-2\sqrt{C_y})-\\&\sum_{y~\in~ A_2\backslash A_3}\mu(y)\xi_ye^{-E_22^{-\frac{n\epsilon}{2}}}(1-2\sqrt{C_y})\\
&\leq 1-\sum_{y\in A_3}\mu(y)(1-D_2e^{d_2n})e^{-E_22^{-\frac{n\epsilon}{2}}}(1-2\sqrt{D_o}e^{-\frac{\Gamma n}{2}})~(d)\\
&=1-\mu(A_3)(1-D_2e^{-d_2n})e^{-E_22^{-\frac{n\epsilon}{2}}}(1-2\sqrt{D_o}e^{-\frac{\Gamma n}{2}}) \\
&\leq 1-\Big[e^{-E_22^{-\frac{n\epsilon}{2}}}(1- (D_1+1)e^{-\min\{d_1,k(\epsilon)\}n})\\&(1-D_2e^{-d_2n})(1-2\sqrt{D_o}e^{-\frac{\Gamma n}{2}})\Big] ~(e)\\
&\leq 1-(1-C'e^{-u(\epsilon)n})e^{-E_22^{-\frac{n\epsilon}{2}}}
\end{split}
\end{equation}
Here, $(d)$ follows from (\ref{xi}) and (\ref{exp_phi}) and ignoring the negative contribution made by the sequences in the set $A_2\backslash A_3$. $(e)$ follows from Lemma \ref{L1}. $C' > 0$ (constant) and $u(\epsilon)$ (positive valued function $\forall~ \epsilon > 0$ and 0 if $\epsilon = 0$) are obtained after simplification of $(e)$. Now, using extended mean value theorem for the function $e^{-z}$, $$e^{-E_22^{-\frac{n\epsilon}{2}}} = 1-E_22^{-\frac{n\epsilon}{2}}+ \frac{e^{-c}}{2}E_2^22^{-n\epsilon}.$$ Here, $c~\in~(0,E_22^{-\frac{n\epsilon}{2}})$. Therefore, we have
\begin{equation}\label{sub_2}
e^{-E_22^{-\frac{n\epsilon}{2}}} \geq 1-E_22^{-\frac{n\epsilon}{2}}.
\end{equation}
Hence, using (\ref{sub_2}) in (\ref{final_r2_sub}), we get
\begin{equation}\label{final_r2}
\mu(\frac{\log R_n(X)}{n} < H -\epsilon) < e^{-g(\epsilon)n}.
\end{equation}
where $g(\epsilon)$ is a positive valued function $\forall ~\epsilon~>~0$ and $g(0) = 0$.
This completes the proof of Theorem \ref{LDP2}.\\
{\it Proof of Corollary \ref{CO43}:}
Using Observation \ref{Obs41}, we have $$L_{2^{n(H-\epsilon)}}(x) > n \Rightarrow R_n(x) \leq 2^{n(H-\epsilon)}$$
$$\Rightarrow \mu(L_{2^{n(H-\epsilon)}}(X) > n)\leq \mu(R_n(X) \leq 2^{n(H-\epsilon)}) < e^{-g(\epsilon)n}.$$
$\forall ~n \geq N'(\epsilon)$. Now, letting $m = 2^{n(H-\epsilon)}$, we have
\begin{equation}
\begin{split}
\mu(L_{m}(X) > \frac{\log m}{H - \epsilon}) < e^{-g(\epsilon)\frac{\log m}{H - \epsilon}}   ~~~~\forall ~m \geq M'(\epsilon)\\
\Rightarrow \mu(\frac{\log m}{L_m(X)} < H- \epsilon) < e^{-g(\epsilon)\frac{\log m}{H - \epsilon}}.
\end{split}
\end{equation}
\begin{Rem}\label{rem4}
{Note that in the first step in Eq. (\ref{final2}) we have a term $1-  \mu(\frac{\log R_n(X)}{n} > H-\epsilon) = \mu(R_n(X) \leq 2^{n(H-\epsilon)})$. Further, in the proof of Theorem \ref{LDP2}, the bound $e^{-g(\epsilon)n}$ is obtained on this term. Hence, there is no ambiguity in using the exponential bound obtained in Theorem \ref{LDP2} on $\mu(R_n(X) \leq 2^{n(H-\epsilon)})$ above in the proof of corollary \ref{CO43}.
}\end{Rem}

\section{{Estimator for Entropy}}\label{c46}
Motivated by experimental results on estimators based on match lengths given in \cite{Suhov98}, we propose an estimator based on \emph{recurrence times} as given below:\\
{\it  Estimator:}
Consider $R_{n,i}(X) = R_n(T^iX).$ \\Define:
\begin{equation}\label{estimator1}
J_n(X) = \frac{1}{Q(n)}\sum_{i = 1}^{Q(n)} \frac{\log R_{n,i}(X)}{n} .
\end{equation}
Here, $Q(n)$ is a function of $n$.
\begin{prop}\label{Prop1}
{If $Q(n)$ is of the polynomial order, then for processes which are exponentially $\phi$-mixing and have exponential rates for entropy,  $$\lim_{n \rightarrow \infty} J_n(X) = H  ~~~~~~~a.s. $$
With $J_n(X)$ satisfying large deviation property.
}\end{prop}
{\it Proof of Proposition \ref{Prop1}:}
\begin{equation}\label{RHS_Estimator}
\begin{split}
\mu(J_n(X) > H+ \epsilon ) &= \mu(\frac{1}{Q(n)}\sum_{i=1}^{Q(n)}\frac{\log R_{n,i}(X)}{n} > H + \epsilon)\\
                                        &\leq \sum_{i=1}^{Q(n)}\mu(\frac{\log R_{n,i}(X)}{n} > H + \epsilon)~(a)\\
                                        &< \sum_{i=1}^{Q(n)}e^{-f(\epsilon)n}   ~~~~\forall~ n \geq  N(\epsilon) ~(b)\\
                                        &= Q(n)e^{-f(\epsilon)n}.
\end{split}
\end{equation}
Here, step $(a)$ follows from Remark \ref{rem8} given in appendix and step $(b)$ follows from the stationarity of the source $X$ and Theorem \ref{LDP1}.
Similarly,
\begin{equation}\label{LHS_Estimator}
\begin{split}
\mu(J_n(X) < H- \epsilon )  &= \mu(\frac{1}{Q(n)}\sum_{i=1}^{Q(n)}\frac{\log R_{n,i}(X)}{n} < H - \epsilon)\\
                                        &\leq\sum_{i=1}^{Q(n)}\mu(\frac{\log R_{n,i}(X)}{n} < H - \epsilon) ~(a)\\
                                        &< \sum_{i=1}^{Q(n)}e^{-g(\epsilon)n}   ~~~~\forall~ n \geq N'(\epsilon)  ~ (b)\\
                                        &= Q(n)e^{-g(\epsilon)n}.
\end{split}
\end{equation}
Here, step $(a)$ follows from Remark \ref{rem8} and step $(b)$ follows from the stationarity of the source $X$ and Theorem \ref{LDP2}.
Therefore, combining (\ref{RHS_Estimator}) and (\ref{LHS_Estimator}), we have
\begin{equation}\label{Tail_Estimator}
\mu(|J_n(X)- H| > \epsilon ) < 2Q(n)e^{-I(\epsilon)n} ~\forall~ n \geq N''(\epsilon)
\end{equation}
where $N''(\epsilon) = \max\{N(\epsilon), N'(\epsilon)\}$.\\
For $Q(n)$ of polynomial order, we have
\begin{equation}\label{BC_Estimator}
\begin{split}
\sum_{n = 1}^{\infty}\mu(|J_n(X)- H| > \epsilon ) &< \sum_{n=1}^{N''(\epsilon)-1}\mu(|J_n(X)-H| >\epsilon)\\& +\sum_{n=N''(\epsilon)}^{\infty}2Q(n)e^{-I(\epsilon)n}\\ &< N''(\epsilon) + \sum_{n=N''(\epsilon)}^{\infty}2Q(n)e^{-I(\epsilon)n}\\                                                                         &<\infty.
\end{split}
\end{equation}
Hence, by Borel-Cantelli Lemma
\begin{equation}
\lim_{n\rightarrow \infty} J_n(X) = H ~~~~~~~ a.s.
\end{equation}
\begin{Rem}\label{rem5}
{The bounds we establish on convergence rate of estimator $J_n(X)$ are loose, we \emph{conjecture} that our proposed estimator will converge to entropy rate at a faster rate than $2e^{-I(\epsilon)n}$.}\end{Rem}

\section{Conclusion}\label{c47}
In this paper, we stated Theorem \ref{OW2} on match lengths which generalizes the match length result given by Ornstein and Weiss~\cite{Ornstein93} for stationary and ergodic processes. Next, for the class of zero entropy processes, we establish through corollaries \ref{OWLT} and \ref{splOWLT}, the behavior of match length asymptotics for irrational rotation and a general zero entropy case respectively. Further, we imitated the proofs given in~\cite{Ziv94} and~\cite{Bansal} of FSLZ and SWLZ algorithms by choosing a $\mathcal{L}_o$ given by Eq. (\ref{102}) in contrast to \emph{their} choice and showed that these algorithms achieve faster convergence rate of the compression ratio for \emph{zero entropy} sequences as compared to those with positive entropy.

 It will be an interesting problem to look for totally ergodic processes that display the behavior given by Eq. (\ref{101}) where the function $f(n)$ is different from $\log n$. With reference to the discussion on zero entropy transformations with discrete and quasidiscrete spectrum, we carry out in Section \ref{c31}, a careful examination of literature might reveal an example where $f(n)$ can be different from $\log n$. Moreover,  the discussion on entropy dimension $\alpha$ that is carried out in Section 3, where it is stated that for each $\alpha \in [0,1)$, there exists a system of entropy dimension $\alpha$, we conjecture (motivated by the analogy between Shannon-MacMillan-Breiman Theorem and Ornstein and Weiss Theorem for positive entropy sources) that for these systems $f(n) = n^\alpha ~ \forall ~\alpha \in (0,1)$. Also, it will be of interest to determine the class of irrational rotations and partitions $\mathcal{P}$ used to generate the sequences for which compression ratio converges to zero, uniformly.

We have proved the large deviation property for the normalized version of recurrence times for exponentially $\phi$-mixing processes. Further, we have also shown this property to hold for our proposed estimator of entropy based on recurrence times. As a future work in this context, it will be interesting to answer if there are faster rate functions than $f(\epsilon)$ and $g(\epsilon)$, and further if the class of processes for which large deviation principle holds can be extended. Also, it will be of interest to prove if LDP also holds true for waiting times. We have established large deviation principle for recurrence times after assuming exponential rates for entropy, it will be of interest to consider if the converse holds true, i.e., if a process satisfies large deviation property for recurrence times, whether it should have exponential rates for entropy.

One can also conduct experimental or theoretical studies comparing the convergence rates of the estimators based on match length as given in \cite{Suhov98} and those based on recurrence times as proposed in this paper.

\section*{Appendix}
Let $\{X_n\}_{n=-\infty}^{n = \infty}$ be a stationary and ergodic process defined on the space of infinite sequences $(\mathcal{A}_{-\infty}^{\infty}, \sigma, \mu)$. Here $\mathcal{A}$ is a finite set of alphabets, $\sigma$ is the sigma field generated by finite dimensional cylinders and $\mu$ is the probability measure.
For simplicity of notation, we will use $X$ for $\{X_n\}_{n = -\infty}^{n = \infty}$. \\
$X$ is called $\psi$-mixing if
\begin{equation}\label{psi}
\sup_{A~\in~\sigma_{-\infty}^n,~ B~\in~\sigma_{n+l}^\infty}\frac{|\mu(A\cap B)-\mu(A)\mu(B)|}{\mu(A)\mu(B)} \leq \psi(l).
\end{equation}
Here, $\psi(l)$ is a decreasing sequence converging to $0$ and $\sigma_i^j$ denotes the sigma algebra generated by $X_i^j = X_iX_{i+1}....X_j$ and it is called $\phi$-mixing if
\begin{equation}\label{phi}
\sup_{A~\in~\sigma_{-\infty}^n,~ B~\in~\sigma_{n+l}^\infty}\frac{|\mu(A\cap B)-\mu(A)\mu(B)|}{\mu(A)} \leq \phi(l).
\end{equation}
Here, $\phi(l)$ is a decreasing sequence converging to $0$.
\begin{Rem}\label{rem6}
{Note that in step (a) in the proof of Theorem \ref{LDP1}, inequality (\ref{UB}) has been used, however it is important to check if it can be applied. This is verified below:  $$\rho_y = \frac{2\sqrt{C_y}}{(\sqrt{1+C_y}+\sqrt{C_y})\xi_y\mu(y)}~~\forall~ y ~\in~ A_n^{(\delta)}$$
Using lower bounds on $\mu(y)$ and $\xi_y$, we have $$\rho_y \leq \frac{2\sqrt{C_y}~2^{n(H+\delta)}}{(\sqrt{1+C_y}+\sqrt{C_y})E_1}~~\forall~y~\in~A_n^{(\delta)}.$$
Since $C_y \rightarrow 0~ (as ~n\rightarrow\infty)$, for a given $d' > 0$, $\frac{2\sqrt{C_y}}{\sqrt{1+C_y}+\sqrt{C_y}} < d'$ for $n$ large enough. Now, we choose $d'$ such that $ 0<d' < E_1$. Since eventually $\delta$ is chosen to be less than $\epsilon$, we have $$ \rho_y < 2^{n(H+\delta)}<2^{n(H+\epsilon)} ~\forall~ y ~\in ~A_n^{(\delta)}.$$
}\end{Rem}
\begin{Rem}\label{rem7}
{Note that, though we prove Theorem \ref{LDP1} under the restriction of certain mixing conditions and using inequality (\ref{UB}), it can also be proved using Markov Inequality and Kac's Lemma under no restriction of mixing. However, the super exponential behavior shown by first term in the proof of Theorem \ref{LDP1} (see Inequality (\ref{final_r1})) is not evident from this alternative proof for mixing sources considered.}\end{Rem}
\begin{Rem}\label{rem8}
{ Let $Z_1, Z_2,....,Z_m$ be $m$ real valued random variables. Consider the following probability, $\mu(\frac{1}{m}\sum_{i=1}^m Z_i  > r)$ and set $E_i = \{\omega: Z_i(\omega) > r\}$. Now,
\begin{equation*}
\begin{split}
\mu(\frac{1}{m}\sum_{i=1}^m Z_i  > r) &\leq \mu(\cup_{i=1}^m E_i)\\
                                                             &\leq \sum_{i=1}^m \mu(E_i)~~(Union~Bound) \\
                                                             &=\sum_{i=1}^m\mu(Z_i > r).
\end{split}
\end{equation*}
Similarly, by changing `$>$' sign with `$<$' accordingly, it can be proved that
$$\mu(\frac{1}{m}\sum_{i=1}^m Z_i < r) \leq \sum_{i=1}^m\mu(Z_i < r).$$
}\end{Rem}
{\it Proof of Lemma \ref{L1}:}
\begin{equation*}\label{Lemma1}
\begin{split}
\mu(A_1\cup A_2)  &\leq 1 \Rightarrow \mu(A_1)+\mu(A_2)-\mu(A_1\cap A_2) \leq 1\\
                   &\Rightarrow \mu(A_1\cap A_2) \geq \mu(A_1) + \mu(A_2) - 1\\
                   &\Rightarrow \mu(A_1\cap A_2) > 1-p_1e^{-p_2n}+ 1-q_1e^{-q_2n} - 1\\
                   &\Rightarrow \mu(A_1\cap A_2) > 1-(p_1e^{-p_2n}+q_1e^{-q_2n})\\
                   &\Rightarrow \mu(A_1\cap A_2) > 1- (p_1+q_1)e^{-\min\{p_2,q_2\}n}.
\end{split}
\end{equation*}
\section*{Acknowledgement}
The authors wish to thank T. Jacob for giving his thoughts on match lengths for zero entropy processes. The authors would also like to thank the reviewers for their constructive comments which have helped in strengthening the paper. The authors would also like to thank Paul Shields for pointing out reference \cite{Kushnirenko} in the context of zero entropy sources. The authors would also like to thank Serap Savari for pointing out certain references.

\begin{IEEEbiographynophoto}{Siddharth Jain}
is currently a graduate student at California Institute of Technology in the department of Electrical Engineering under the supervision of Prof. Jehoshua (Shuki) Bruck. His research interests include information systems, combinatorics, data compression and applications of information theory in computational biology. Siddharth received Bachelors and Masters degree from Indian Institute of Technology (IIT) Kanpur in 2013. He did his Masters Thesis under the supervision of Prof. R.K. Bansal. He was awarded the proficiency medal at IIT Kanpur for excellent academic performance in Electrical Engineering.
\end{IEEEbiographynophoto}

\begin{IEEEbiographynophoto}{Rakesh K. Bansal}
did his B. Tech. at IIT Kanpur, India in 1978 and completed his M.S.  and Ph.D. in 1983 and 1987 respectively from University of Connecticut, Storrs. All his degrees are in Electrical Engineering. After spending a year at University of Virginia, he joined the faculty at IIT Kanpur, Department of Electrical Engineering in 1988 where he is currently a Professor. His research interests center around universal compression and source coding, sequential detection of a change distribution and robustness. 
\end{IEEEbiographynophoto}
\end{document}